\documentclass[aOAps,fullpaper,twocolumn,footinbib]{revtex4-1}
\usepackage[dvipdfmx]{graphicx}
\usepackage{amsmath}
\usepackage{amsfonts}
\usepackage{amssymb}
\usepackage{braket}
\usepackage{bm}
\usepackage{color}
\usepackage{verbatim}

\begin{document}

\title{$\mathbb{Z}_Q$ Berry Phase for Higher-Order Symmetry-Protected Topological Phases}
\author{Hiromu Araki}
\email{araki@rhodia.ph.tsukuba.ac.jp}
\affiliation{Graduate School of Pure and Applied Sciences, University of Tsukuba, Tsukuba, Ibaraki 305-8571, Japan}
\author{Tomonari Mizoguchi}
\affiliation{Department of Physics, University of Tsukuba, Tsukuba, Ibaraki 305-8571, Japan}
\author{Yasuhiro Hatsugai}
\affiliation{Department of Physics, University of Tsukuba, Tsukuba, Ibaraki 305-8571, Japan}
\begin{abstract}
We propose  the $\mathbb{Z}_Q$ Berry phase as a topological invariant for
higher-order symmetry-protected
topological (HOSPT) phases for  two- and three-dimensional systems. 
It is topologically stable for
electron-electron interactions assuming the gap remains open.
As a concrete example, 
we show that the Berry phase 
is quantized in $\mathbb{Z}_4$ and characterizes the HOSPT phase of the extended 
Benalcazar-Bernevig-Hughes (BBH) model, which contains the next-nearest neighbor 
hopping and the intersite Coulomb interactions. 
In addition, we introduce the $\mathbb{Z}_4$ 
Berry phase for the spin-model-analog of the BBH model,
whose topological invariant has not been found so far.
Furthermore, we demonstrate the Berry phase is quantized in $\mathbb{Z}_4$
for the three-dimensional version of the BBH model.
We also confirm the bulk-corner correspondence between the $\mathbb{Z}_4$ Berry phase and 
the corner states in the HOSPT phases.
\end{abstract}

\maketitle

\textit{Introduction.-}
Topological phases of matter, distinct from the conventional phases 
in that they are not characterized by the local order parameter but by the topological order parameter, 
have been one of the central topics of the condensed matter physics. 
Even ten years after the celebrated ten-fold-way classification~\cite{PhysRevB.78.195125,doi:10.1063/1.3149495,1367-2630-12-6-065010} of the topological insulators and superconductors~\cite{RevModPhys.82.3045,RevModPhys.83.1057}, 
the notion of topologically nontrivial states in non-interacting fermions has greatly extended its scope
by incorporating the crystalline symmetries~\cite{PhysRevLett.106.106802, Hsieh2012, Slager2012, doi:10.1146/annurev-conmatphys-031214-014501, Wang2016, PhysRevB.95.224514, PhysRevX.7.041069, PhysRevLett.120.266401, PhysRevLett.121.126402, 
Slager2018}.
It was further revealed that short-range entangled quantum many-body states 
can also host topologically nontrivial state protected by symmetries, 
and they are now unified by the notion of symmetry protected topological phases 
(SPT phases)~\cite{PhysRevB.81.064439,PhysRevB.85.075125,doi:10.1146/annurev-conmatphys-031214-014740,RevModPhys.89.041004}.

Recently, a novel topological states of matter associated with crystalline symmetries, called a higher-order topological insulator (HOTI), were proposed~\cite{Volovik2010, PhysRevB.92.085126, PhysRevB.95.165443,Hayashi2018, Schindlereaat0346,PhysRevB.96.245115,Benalcazar61,PhysRevLett.120.026801, PhysRevB.97.241405,Xu2017,Hayashi2019,PhysRevB.99.085426}. 
This state has topologically protected boundary states 
with co-dimension larger than one, e.g., corner states in two- and three-dimensional systems.
Together with these theoretical developments,
experimental realization of the HOTIs has also been intensively pursued
both in solid-state systems~\cite{Schindler2018} and artificial materials~\cite{Garcia2018,Imhof2018,PhysRevB.98.205147,NatMaterXue2019,Ota:19}.

So far, to identify the HOTI phase, several topological invariants have been proposed,
such as the nested Wilson loop~\cite{PhysRevB.96.245115,Xu2017}, 
the quantized Wannier center~\cite{PhysRevLett.120.026801}, 
the entanglement polarization/entropy~\cite{PhysRevB.98.035147,Wang_2018}
and the multiple moment~\cite{Kang2018}.
The $K$-theoretic classification was also proposed~\cite{PhysRevB.99.085127}.
Yet, not many examples are known to be applicable to the quantum many-body analog of the HOTI phase, 
or the higher-order symmetry protected topological phase (HOSPT phase)~\cite{PhysRevB.98.235102,PhysRevB.99.235132,arXiv:1809.07325},
which includes not only interacting fermion systems but also spin (bosonic) systems.
It is therefore highly desirable to find a topological invariant
which can be used to identify the HOTI and HOSPT phases, ranging from non-interacting fermion systems, to bosonic/fermionic many-body systems.

In this Letter, we propose that 
the quantized Berry phase with respect 
to the local twist of the Hamiltonian 
characterizes the HOTI and HOSPT phases. 
In the literature, the quantized Berry phase has been used for characterizing various SPT phases, 
including both non-interacting systems and quantum many-body 
systems~\cite{doi:10.1143/JPSJ.75.123601, 1367-2630-12-6-065004, PhysRevB.77.094431, Hatsugai_2007, 
PhysRevE.87.021301, PhysRevB.94.205112, PhysRevLett.120.247202, PhysRevB.98.195127, 
doi:10.7566/JPSJ.88.045001, PhysRevB.100.014438}.
The key observation in those examples is that, 
finite Berry phase indicates 
that the ground state is adiabatically connected with 
the ``irreducible cluster state'',
which cannot be decomposed in to the smaller elements under the symmetries which protect the topological phases.  
Here, we demonstrate that the HOSPT state can also be connected to 
the irreducible cluster state, and that the characteristic higher-order boundary states 
can be obtained by ``amputating'' the clusters at the boundary. 
As such, the quantized Berry phase serves as a topological invariant for the HOSPT phase, similarly to the conventional SPT phases. 

As a concrete example, 
we employ the seminal model of the HOTI introduced 
by Benalcazar-Bernevig-Hughes (BBH)~\cite{PhysRevB.96.245115} 
with the additional next-nearest-neighbor (NNN) hopping term.
We show that four-fold rotational ($C_4$) symmetry 
gives rise to the $\mathbb{Z}_4$ Berry phase.
We then extend our target to the many-body analogs of the BBH model, 
namely, the BBH model with the intersite repulsive interaction and the spin-model analog of the BBH model, 
which are the platforms of the HOSPT phases. 
In both of these two models, the correspondence between the $\mathbb{Z}_4$ Berry phase and the gapless corner excitation 
for the finite system is confirmed, which clearly 
demonstrates that the quantized Berry phase characterizes the HOSPT phases beyond the non-interacting fermion systems. 
Finally, the application 
of the present formalism to the three-dimensional BBH model (3D BBH model) is discussed.

\begin{figure}[t]
 \centering
 \includegraphics[width=1\linewidth]{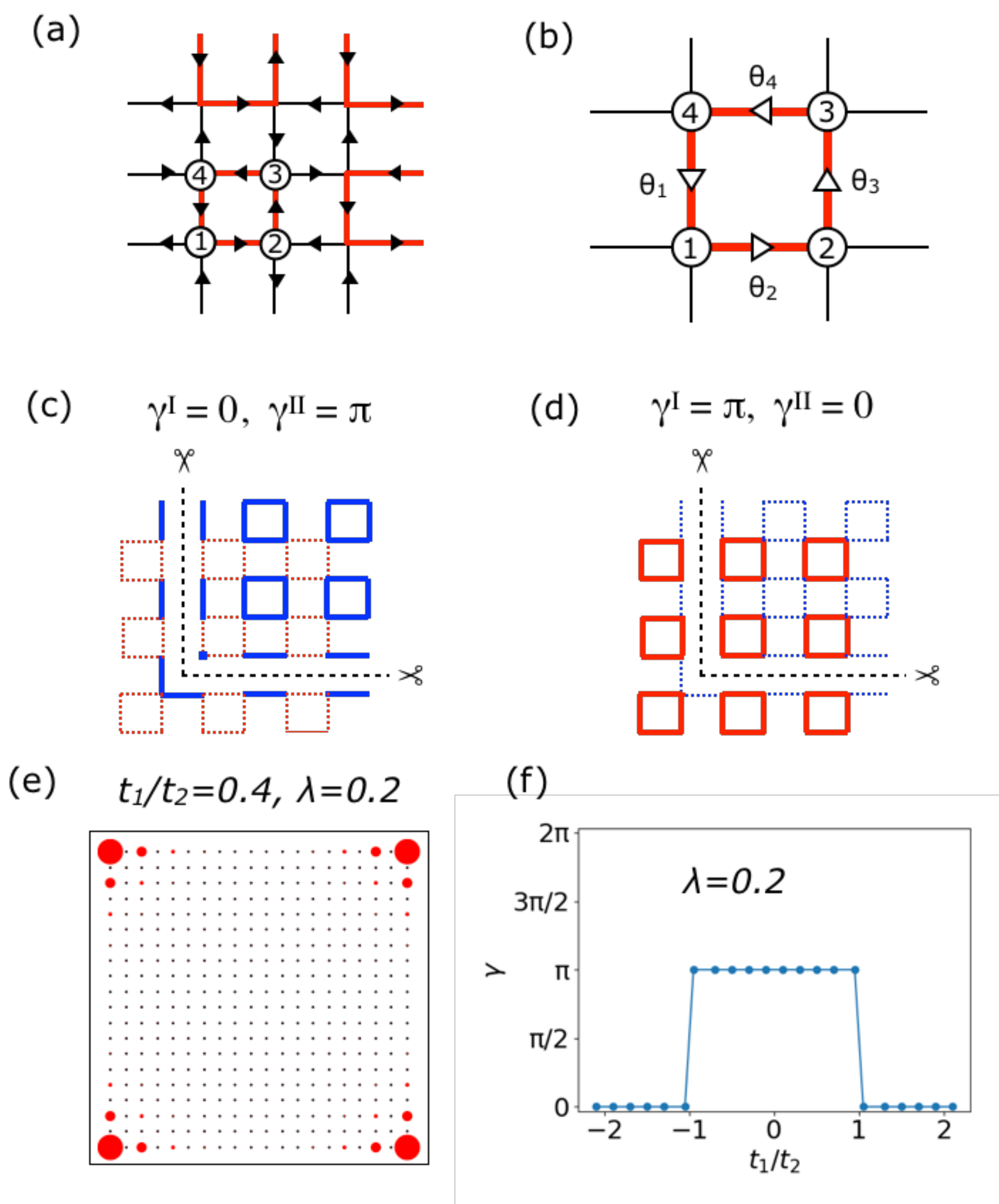}
 \caption{
 (a) Hopping terms of the square lattice model is shown. The amplitude of the hopping 
 on red (blue) lines are $t_1$ ($t_2$). 
 The phases of the hopping are $e^{-i\pi/4}$ along the arrows.
 (b) The twist parameters $\theta_1,...,\theta_4$ for the $\mathbb{Z}_4$ Berry phase are shown.
 (c)-(d) Schematic picture about the correspondence between the corner state and the Berry phases. 
 The red (blue) plaquettes are the type-I (II) plaquettes. 
 Thick (dashed) lines denote the strong (weak) bonds.
 In (c)/(d), the plaquettes with strong/weak bonds are cut to make a corner.
 In (c), the corresponding Berry phase for strong/weak plaquettes are $\gamma^{\rm I}=0/\gamma^{\rm II}=\pi$ 
 and the corner state appears, 
 while in (d), $\gamma^{\rm I}=\pi/\gamma^{\rm II}=0$ and no corner states appear.
 (e) The total density of four corner states is plotted for the system with $(t_1/t_2 = 0.4, \lambda=0.2)$.
 The system size is $20\times 20$. 
 (f) The Berry phase for the model with $\lambda=0.2$ against $t_1/t_2$. The non-trivial Berry phase 
 corresponds to the HOTI phase.
 }
 \label{fig:fig1}
\end{figure}

\textit{$\mathbb{Z}_4$ Berry phase for non-interacting fermions.-}
The Hamiltonian for the extended BBH model reads
$\mathcal{H}_{0} = \mathcal{H}_{NN} + \mathcal{H}_{NNN}$, where $\mathcal{H}_{NN} = -\sum_{\braket{ij}} t_{ij} 
e^{i \alpha_{i,j} } c_i^\dagger c_j$, and
$\mathcal{H}_{NNN} = -\lambda \sum_{\braket{\braket{ij}}} u_{ij} c_i^\dagger c_j$.
 $\mathcal{H}_{NN}$ represents the NN-hopping term, and
 $t_{ij} = t_1$ $(t_2)$ for bonds colored in red (blue) in Fig. \ref{fig:fig1}(a).
The phase factor $e^{i \alpha_{i,j} }$ is chosen such that 
the $\pi$-flux is inserted to every square plaquette, 
which is essential to obtain the bulk energy gap~\cite{Hayashi2018, Hayashi2019}. 
We set $\alpha_{i,j} = \pi/4$ along the arrows shown in Fig.~\ref{fig:fig1}(a)
to explicitly represent the $C_4$ symmetry.
Note that $\mathcal{H}_{NN}$ is equivalent to the original form shown in Ref.~\onlinecite{PhysRevB.96.245115}, 
which seemingly lacks the $C_4$ symmetry,
under the gauge transformation.
For convenience, we label the square plaquettes in three types:
type-I, where all bonds have the hopping amplitude $t_1$,
type-II, where all bonds have the hopping amplitude $t_2$,
and type-III, where two of four bonds have the hopping amplitude $t_1$ and the rest have $t_2$.
Then, in the NNN-hopping term $\mathcal{H}_{NNN}$, 
$u_{ij}$  is set according to the type of the plaquette to which the NNN bond belong, namely,
$u_{ij} = t_1, t_2$, and $(t_1+t_2)/2$ if the bond $(i,j)$ 
is in the type-I, type-II and type-III plaquettes, respectively.
The parameter $\lambda$ in $\mathcal{H}_{NNN}$ controls the ratio between $\mathcal{H}_{NN}$ and $\mathcal{H}_{NNN}$.
We emphasize that the model with finite $\lambda$ has the $C_4$ symmetry but broken chiral symmetry.
In the following, if not mentioned otherwise, we consider the case of half-filling.

Now, let us define the Berry phase with respect to the local twist of the Hamiltonian~\cite{0295-5075-95-2-20003}.
The local twist is introduced in the following manner. 
To begin with, we rewritten $\mathcal{H}_0$ as
$\mathcal{H}_0 = \sum_{\eta = \mathrm{I, II} } \sum_{P \in \mathrm{type-}\eta} h_{P}$, 
where $h_{P}$ is the Hamiltonian of the plaquette $P$.
We choose one of the square plaquettes $P_0$ which belongs to either type-I or type-II.
We then 
modify $h_{P_0} $ in such a way that 
$h_{P_0} (\bm{\Theta}) = - \sum_{\langle ij \rangle \in P_0} t_{ij} e^{i \alpha_{i,j}} \tilde{c}^\dagger_i \tilde{c}_j
-\lambda \sum_{\braket{\braket{ij}} \in P_0} u_{ij}\tilde{c}_i^\dagger \tilde{c}_j$,
where 
$\tilde c_j  := e^{i \varphi_j} c_j $
with $\varphi_j = \sum_{q = 1}^{j}\theta_q$ for $j = 1,2,3,4$
and $\varphi_4=0$. 
The parameter space $\bm{\Theta}$ is defined by three independent parameters
$(\theta_1, \theta_2, \theta_3)$ and $\theta_4=-\sum_{j=1}^{3}\theta_j$.
Note that the Hamiltonians on all the other plaquettes are not changed.
We write the total Hamiltonian with the twist as 
$\mathcal{H} (\bm{\Theta}) := h_{P_0} (\bm{\Theta}) + \sum_{P \neq P_0} h_{P}$. 
We define trajectories $L_j (j=1,2,3,4)$ in the parameter space:
$\bm{E}_{j-1} \rightarrow \bm{G} \rightarrow \bm{E}_j$
where 
$\bm{E}_1 = (2\pi, 0, 0)$, $\bm{E}_2 = (0, 2\pi, 0)$, $\bm{E}_3 = (0, 0, 2\pi)$,
$\bm{E}_0 = \bm{E}_4 = (0, 0, 0)$, 
and $\bm{G} = 1/4 \sum_{j=1}^{4}\bm{E}_j$.
The Berry phase for the parameter space 
is defined as a contour integral of the Berry connection, 
$\bm{A} (\bm{\Theta}) =\braket{\Psi( \bm{\Theta}) | \frac{\partial}{\partial \bm{\Theta}} | \Psi( \bm{\Theta})}$, 
along the path $L_j$, 
$\gamma^{\eta}_{j}= - i \oint_{L_{j}} d \bm{\bm{\Theta}} \cdot  \bm{A}(\bm{\Theta})$,
where $\ket {\Psi(\bm{\Theta})}$ represents the many-body ground state for 
$\mathcal{H} (\bm{\Theta})$. 
For non-interacting fermions, $\ket {\Psi(\bm{\Theta})}$ can be obtained by occupying all the single-particle states having negative energy. 

The Berry phase for the present model is quantized in 
$\mathbb{Z}_4$ because of the following reason:  
Firstly, due to the cancellation of the trajectories in each path $L_j$,
we have
\begin{equation}
\sum_{j=1}^{4} \gamma^\eta_{j} \equiv 4 \gamma \equiv 0~ \bmod ~2 \pi.
\end{equation}
Secondly, the $C_4$ symmetry enforces 
\begin{equation}
 \gamma^\eta_{1} \equiv \gamma^\eta_{2} \equiv \gamma^\eta_{3}  \equiv \gamma^\eta_{4} \equiv \gamma^{\eta} ~ \bmod ~2 \pi.
\end{equation}
Combining these two equations, we obtain 
$\gamma^\eta_j \equiv 2 \pi \frac{n}{4}~ \bmod 2 \pi, ~ n \in \mathbb{Z}$
for $j=1,2,3,4$.
In the following, we abbreviate $\gamma^\eta_j$ as $\gamma^\eta$.
\begin{figure*}[t]
 \centering
 \includegraphics[width=1.0\linewidth]{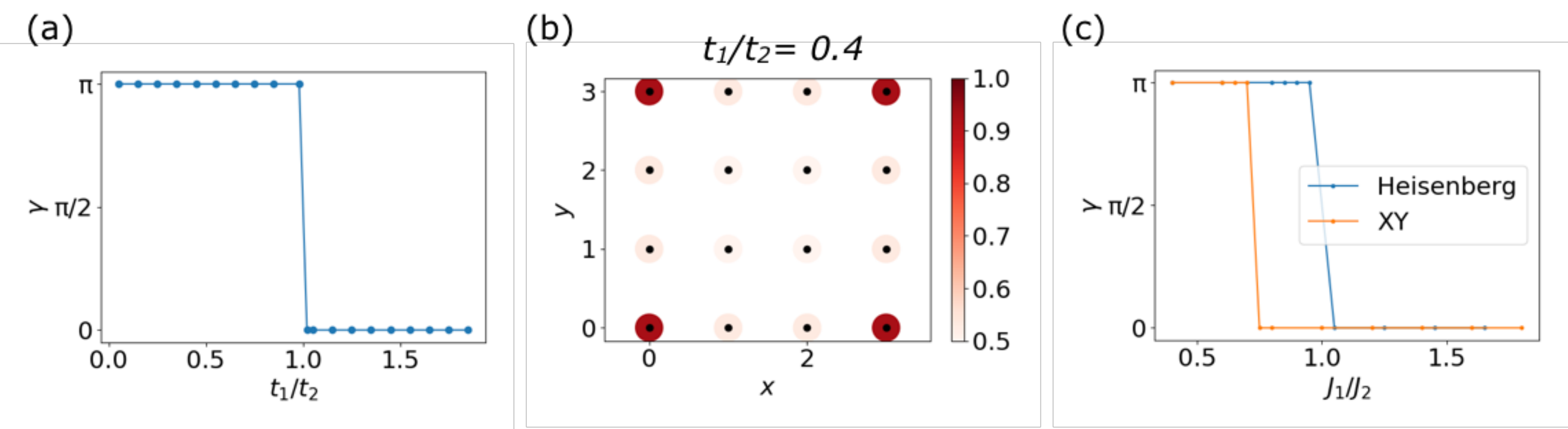}
 \caption{
 (a) The $\mathbb{Z}_4$ Berry phase for the interacting fermion model. The Berry phases are nicely 
 quantized even with the interactions.
 The non-trivial phase is adiabatically connected to the non-interacting HOTI phase.
 (b) The occupation number at representative sites indicated in the inset, where $(L\times L)/2+2$ particles are
 filled. The results are obtained for the systems with $L=4$.
(c) The $\mathbb{Z}_4$ Berry phases for 
 the BBH-type spin model.}
 \label{fig:fig3}
\end{figure*}

Physical consequences of the nontrivial Berry phase can be well-illustrated 
by considering two ``decoupled'' limits, namely, $t_1=0, \hspace{.5mm} t_2 \neq 0$
and  $t_1\neq 0, \hspace{.5mm} t_2  =0$.
In the former limit, the Hamiltonian is given by $\mathcal{H} = \sum_{P \in \mathrm{type-II}} h_{P}$,
thus the ground state is nothing but the product state of the plaquette state,
$\ket {\Psi_{0}^{\mathrm{II}}} = \prod_{P  \in \mathrm {type-II} } \left( \psi^\dagger_{P,2}  \psi^\dagger_{P,1} \ket{0}_P \right)$,
where $\psi^\dagger_{P,1}$ and $\psi^\dagger_{P,2}$ are the lowest and the second-lowest energy states of 
$h_{P}$, respectively, and $\ket{0}_{P}$ is the vacuum of $P$.
We refer to $\ket {\Psi_{0}^{\mathrm{II}}}$ as the type-II plaquette state.
Then, one can show that
$\gamma^{\rm I} = 0$ since the Hamiltonian on type-I plaquettes is switched off in this limit, 
and that $\gamma^{\rm II} = \pi = 2 \pi \cdot \frac{1}{2}$, 
reflecting the fact that the Berry phase for the decoupled cluster corresponds to the filling factor multiplied by $2\pi$ 
(see Supplemental Material for details~\cite{SM}).
Now, let us switch on $t_1$. 
As far as $|t_1| < |t_2|$ is satisfied, the bulk band gap does not close upon increasing $t_1$,
thus the Berry phase does not change even for finite $t_1$. 
This implies that the ground state for $|t_1| < |t_2|$ is adiabatically connected to the type-II plaquette 
state [Fig. \ref{fig:fig1} (c)].
One can also show that, if we start from the latter limit, 
i.e. $t_1\neq 0, \hspace{.5mm} t_2  =0$, 
the ground state is adiabatically connected to the type-I plaquette state,  $\ket {\Psi_{0}^{\mathrm{I}}}$ 
as far as $|t_1| > |t_2|$ is satisfied. Thus, the ground state has the Berry phase $\gamma^{\rm I}=\pi$ [Fig. \ref{fig:fig1}(d)].
We emphasize that the plaquette states discussed 
above are minimally decoupled states 
connected to the ground state of $\mathcal{H}_0$. Since the plaquette states cannot be adiabatically connected 
to the atomic insulator, they are the ``reference states'' of the HOSPT phase in the present model.

Having the decoupled picture at hand, the boundary states on the finite systems is naturally inferred,
namely, if the ground state is connected to the type-$\eta$ plaquette state
and the type-$\eta$ plaquette is cut off a the corner, 
there has to be a zero energy state at the corner which does not belong to any type-$\eta$ plaquette.
We demonstrate this picture for the model $\mathcal{H}_0$. 
Consider the system under the open boundary condition, whose corner configuration is chosen such that the type-II plaquette is cut off. 
In this model, the exact corner states can be constructed 
for $|t_1/t_2| < 1$ at any $\lambda$ [Fig.~\ref{fig:fig1}(e)]~\cite{SM}, whereas the corner state does not exist for
$|t_1/t_2| > 1$, meaning that the phase transition form the HOTI phase to the trivial phase occurs at $t_1/t_2 = \pm 1$. 
Turing to the system under periodic boundary conditions, 
the $\mathbb{Z}_4$ Berry phase $\gamma^{\mathrm{II}}$ becomes nontrivial for $|t_1/t_2| < 1$ [Fig.~\ref{fig:fig1}(f)],
which completely coincides with the HOTI phase.
Note that there is a relation between $\gamma^{\rm I}$ and $\gamma^{\rm II}$ such 
that $\gamma^{\rm I}(t_1/t_2) = \gamma^{\rm II}(t_2/t_1)$, indicating the duality between type-I and type-II plaquettes.
Considering these, we conclude that the bulk-boundary correspondence 
between the $\mathbb{Z}_4$ Berry phase and the zero-energy corner states of the HOTI phase holds, 
thus the $\mathbb{Z}_4$ Berry phase indeed serves as a topological invariant for the HOTI phase. 

\textit{Interacting fermions.-} 
We now turn to the results of the many-body systems.
The many-body eigenvalues and eigenstates 
are calculated by the exact diagonalization using the lattice-model solver 
$\mathcal{H}\Phi$~\cite{KAWAMURA2017180}.
We begin with the BBH model with the NN repulsive interaction
$\mathcal{H} = \mathcal{H}_0 + \mathcal{H}_{\rm int}$
with $\mathcal{H}_{\rm int} = V \sum_{\braket{ij}} \hat{n}_i \hat{n}_j$ where $\hat{n}_i = c^\dagger_i c_i$ 
represents the density operator.
We employ a finite system with $N = L\times L $ sites, and consider the half-filled case. 
For simplicity, we set $\lambda =0$ in the following.

\begin{figure}[b]
 \centering
 \includegraphics[width=0.95\linewidth]{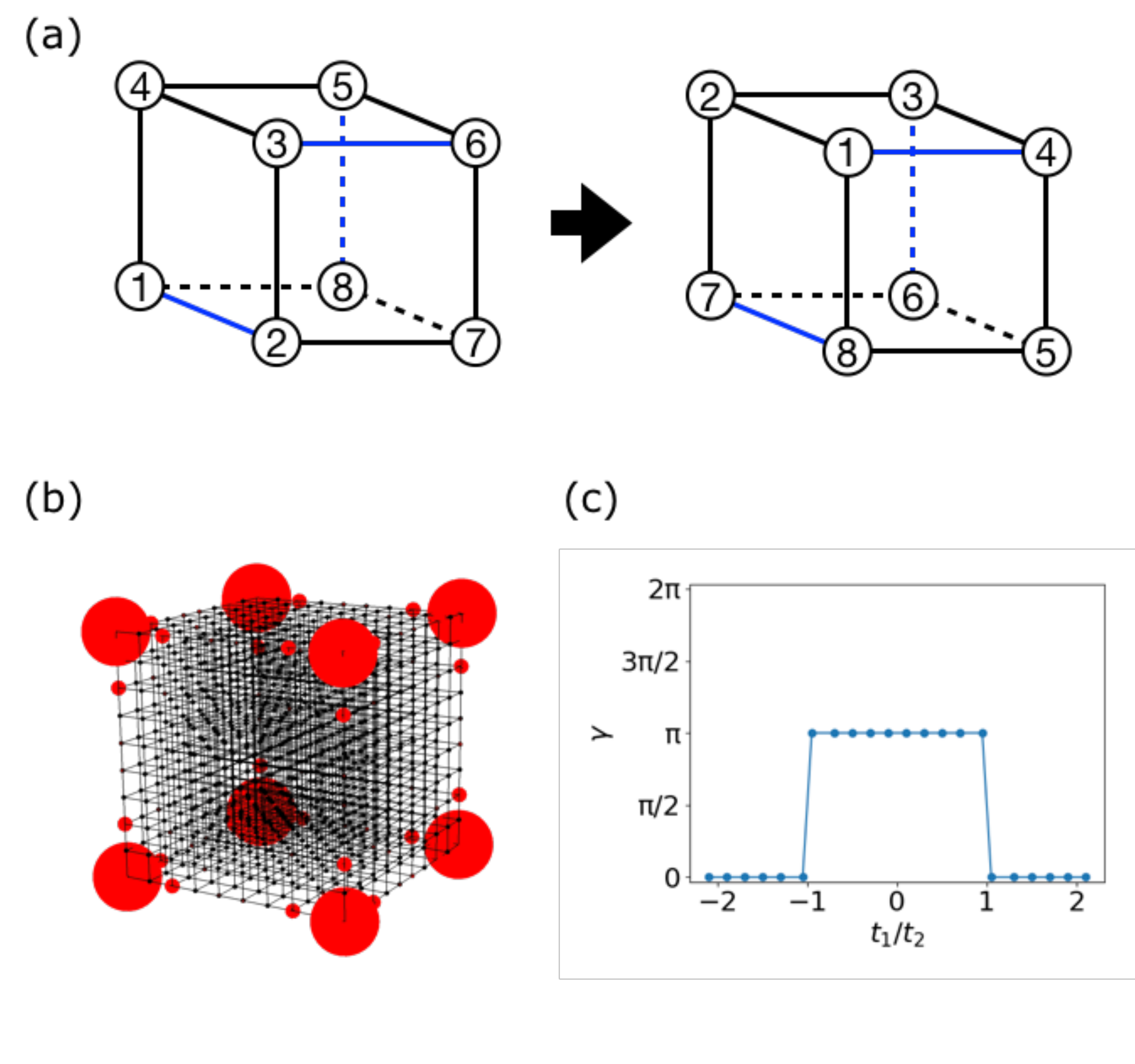}
 \caption{
 (a) The unit cell of the 3DBBH model is shown. 
 On the blue line, the hopping term have the phase $e^{-i \pi}=-1$. Then the model has a $\pi$-flux for each plane.
 The $\mathbb{Z}_4$ operation change the basis of the unit cell from left to right.
 (b) The total density of eight corner states is plotted with $t_1/t_2=0.1$. The system size is $10 \times 10$.
 (c) The Berry phase for the model against $t_1/t_2$.
 }
 \label{fig:fig2}
\end{figure}

We have numerically confirmed that the ground state is gapped for $V \geq 0$ under the periodic boundary condition.
Note, however, that the quantum phase transition to the charge density wave will occur at $V =V_c$ upon increasing $V$,
if we consider the thermodynamic limit ($ L \rightarrow \infty$).
Nevertheless, one can expect that the $V_c$ is larger than the bulk band gap, thus
the following result will be valid if $V$ is smaller than the bulk band gap. 
In Fig.~\ref{fig:fig2}(a), we plot the Berry phase for $V=0.4$ as a function of $t_1/t_2$.
Clearly, the Berry phases are quantized and the topological phase transition occurs upon changing $t_1/t_2$.

Similarly to the non-interacting-fermion analog, $\pi$-Berry phase indicates the topologically nontrivial state, or the HOSPT phase.
To confirm this, we examine the spacial profile of the particle distribution of the charge excitation
under the open boundary condition in both of two directions.
To be more concrete, we increase the number of particles from
$\frac{L \times L}{2}$ to $\frac{L \times L}{2} + 2$, and investigate the occupation number of each site.
If the low-energy excitation is localized at the corners, 
one can expect that the occupation number 
become $1$ only at the corners while it remain to be $1/2$ in the bulk,
which becomes a hallmark of the HO topological phase.

The result is shown in Fig.~\ref{fig:fig2}(b).
We see that the occupation numbers at corner sites are enhanced, while the occupation numbers at bulk sites remain $1/2$.
This means that gapless excitations that are reminiscent of the corner zero mode of the 
HOTI in the non-interacting case are localized at the corner, as expected.
We then conclude that the HOSPT phase which is characterized by the gapless corner excitation 
exists for the interacting BBH model, and the $\mathbb{Z}_4$ Berry phase serves as a topological invariant 
as is in the non-interacting case. 

\textit{Spin model.-}
As the second example of the quantum many-body system with the HOSPT phase,
we study the spin-model analog of the BBH model~\cite{PhysRevB.99.235132},
$\mathcal{H}_{\rm spin} = \sum_{\braket{i, j}} J_{ij} \left[\frac{1}{2} \left(S^+_i S^-_j + S^-_i S^+_j \right) + \Delta S^z_i S^z_j\right]$,
where $\bm{S}_i$ is the spin operator of $S=1/2$ at the site $i$, 
$J_{ij} = J_1, J_2$ are the exchange parameters on the NN bonds and  
$\Delta$ is the Ising anisotropy; $\Delta = 0$ and $\Delta = 1$
correspond to the quantum XY model and the Heisenberg model, respectively. 
The spacial configuration of $J_{ij}$ is the same as that for 
the non-interacting-fermion analog, 
obtained by the replacement $t_1 \rightarrow J_1$ and $t_2 \rightarrow J_2$.
The existence of the corner modes in this model is discussed by using Jordan-Wigner transformation 
in Ref.~\onlinecite{PhysRevB.99.235132}.

To define the Berry phase, 
we again decomposed the Hamiltonian into the sum over the plaquettes 
and introduce the twist parameters as 
$S^-_j \rightarrow e^{i \varphi_j}S^-_j$ and $S^+_j \rightarrow e^{-i \varphi_j}S^+_j$,
on one of the plaquettes which belongs to either type-I or type-II. 
Then, the Berry phase is given by the same form as that of fermion systems.
Figure \ref{fig:fig3}(c) shows that the Berry phase $\gamma^{\rm II}$. 
Again, the clear change of $\gamma^{\rm II}$ is seen, and
$\gamma^{\eta} = \pi$ indicates that the state can be 
adiabatically connected to the irreducible cluster state on the type-$\eta$ plaquette,
i.e., the state is in the HOSPT phase. 
Interestingly, in contrast to the fermionic systems, 
the transition point for the XY model is deviated from $J_1/J_2 = 1$, meaning that
there is an intermediate phase where both of the Berry phases are equal to zero, 
which cannot be connected to neither of the decoupled cluster states.
This can be artifact arising from the finite size effect, and identifying 
the nature of this phase requires further studies. 

\textit{3D BBH model.-}
Finally, we apply the Berry phase analysis to the 3D BBH model. 
The Hamiltonian for the 3D BBH model reads~\cite{PhysRevB.96.245115,Benalcazar61}
$\mathcal{H}^{3D}_{0} = -\sum_{\braket{ij}} e^{-i \alpha_{i,j}} t_{ij} c_i^\dagger c_j$,
where $t_{ij} = t_1$ for bonds in unit cells and otherwise $t_{ij} = t_2$. 
The phases of the hopping are $\alpha = \pi (0)$ for bonds along the blue (black) lines in 
Fig. \ref{fig:fig2} (a), hence all the surfaces have $\pi$-flux per an unit surface.
The model has the HOTI phase when $t_1/t_2 < 1$. Figure \ref{fig:fig2} (b) shows the 
total density plot of the corner states with $t_1/t_2 = 0.1$.

To define the Berry phase, we again introduce the twist in the Hamiltonian as
$\tilde c_j  := e^{i \varphi_j} c_j $
with $\varphi_j = \sum_{q = 1}^{j}\theta_q$ for $j = 1,2,\cdots,8$
and $\varphi_8=0$. Then we have seven independent parameters 
$\bm{\Theta} = (\theta_1, \cdots, \theta_7)$.
As shown in Fig. \ref{fig:fig2} (a), the Hamiltonian is invariant under $(\mathbb{Z}_8)^2$ 
symmetry.
In the parameter space, we define trajectories $L_j (j=1,2,\cdots,8)$:
$\bm{E}_{j-1} \rightarrow \bm{G} \rightarrow \bm{E}_{j}$
where 
$\bm{E}_1 = (2\pi, 0, \cdots, 0)$, $\cdots$, $\bm{E}_7 = (0, 0, \cdots, 2\pi)$,
$\bm{E}_0 = \bm{E}_8 = (0, 0, \cdots, 0)$, 
and $\bm{G} = 1/8 \sum_{j=1}^{8}\bm{E}_j$.
Due to the $(\mathbb{Z}_8)^2$ symmetry, 
the Berry phase $\gamma_{j}= - i \oint_{L_{j}} d \bm{\bm{\Theta}} \cdot  \bm{A}(\bm{\Theta})$
have an equation
$\gamma_1 + \gamma_2 \equiv \gamma_3 + \gamma_4 \equiv \gamma_5 + \gamma_6 
\equiv \gamma_7 + \gamma_8 \equiv \gamma ~\bmod ~ 2 \pi$.
Then the Berry phase $\gamma$ is quantized in $\mathbb{Z}_4$.
Figure \ref{fig:fig2} (c) shows the Berry phase $\gamma$ against $t_1/t_2$, which clearly shows
the non-trivial Berry phase corresponds to the HOTI phase in $t_1/t_2 < 1$.

\textit{Summary and discussions.-}
We have demonstrated that the $\mathbb{Z}_4$ Berry phase characterizes
the HOSPT phases in the $C_4$ symmetric square lattice models.
The key idea comes from the fact that the ground states are
adiabatically connected to the product states of the decoupled irreducible clusters.
The bulk-corner correspondence in these systems is then naturally understood as 
a consequence that the boundary cuts the clusters 
such that the isolated site(s) appears. 
Numerical evidences of the above are presented for
the free-fermion BBH model, the BBH model with the NN interaction, and the spin-model-analog of the BBH model.
Further, we have shown that the quantized Berry phase characterizes the HOSPT phase in the 3D BBH model as well. 

In this Letter, we have focused on the BBH-type models with the $C_4$ symmetry, and  it is worth noting the protecting symmetries of the BBH model. It was argued that two mirror symmetries are enough to protect the HOTI phase~\cite{Benalcazar61,Kang2018}, 
instead of the $C_4$ symmetry.
If the $C_4$ symmetry is broken while two mirror symmetries are kept, 
the ground state can be adiabatically connected to the valence-bond-solid state on the strong bonds, 
which can be captured by the conventional $\mathbb{Z}_2$ Berry phase~\cite{doi:10.1143/JPSJ.75.123601,Hatsugai_2007,arXiv:1809.08248}. 
However, at the $C_4$-symmetric point, the valence-bond state is not the irreducible cluster state since it does not respect the $C_4$ symmetry. 
Consequently, the $\mathbb{Z}_2$ Berry phase becomes ill-defined, and we need to use the $\mathbb{Z}_4$ Berry 
phase, which we introduced in this Letter.

It is also worth noting that  
the quantized Berry phase can be straightforwardly applied to systems with the $\mathbb{Z}_Q$ symmetry or the $C_Q$ symmetry. 
The examples include
the $\mathbb{Z}_3$ Berry phase for $C_3$ symmetric breathing kagome model~\cite{SM, 0295-5075-95-2-20003,Kudo2019}
and the $\mathbb{Z}_6$ Berry phase for $C_6$ symmetric honeycomb lattice model~\cite{doi:10.7566/JPSJ.88.104703}.
Since various HOTI/HOSPT phases with the $\mathbb{Z}_Q$ or $C_Q$ symmetry have been proposed~\cite{PhysRevB.99.245151},
we believe that the $\mathbb{Z}_Q$ Berry phase is a powerful tool to study such phases.

We would like to thank Tsuneya Yoshida and Koji Kudo for fruitful discussions. 
This work is supported by the JSPS KAKENHI, Grant number JP17H06138 and JP16K13845, 
MEXT, Japan.

%


\pagebreak
\widetext
\begin{center}
\textbf{\large Supplemental material for ``$\mathbb{Z}_Q$ Berry Phase for
Higher-Order Symmetry-Protected Topological Phases''}
\end{center}
\setcounter{equation}{0}
\setcounter{figure}{0}
\setcounter{table}{0}
\setcounter{page}{1}
\makeatletter
\renewcommand{\theequation}{S\arabic{equation}}
\renewcommand{\thefigure}{S\arabic{figure}}
\renewcommand{\bibnumfmt}[1]{[S#1]}

\appendix

\section*{$\mathbb{Z}_Q$ Berry phases}
\subsection{Hyper-pyrochlore lattice models}

Recently, it is found that the breathing kagome model and breathing Pyrochlore model 
have the higher-order topological insulator (HOTI) phase [27, 29].
In the HOTI phases, the models have mid-gap corner states, which is exactly solvable with certain boundary
conditions [28, 31].
In this section,  we show the correspondence between the $\mathbb{Z}_Q$ Berry phases and the HOTI phases
in the breathing kagome model and the breathing pyrochlore model. 
The Su-Schrieffer-Heeger (SSH) model, the breathing kagome model and 
the breathing pyrochlore are regarded as the $d=1, 2, 3$-dimensional breathing hyper-tetrahedron (BHT) model, 
respectively, which have $d+1$ sites in a unit cell. 
For the $d$-dimensional BHT model, the $\mathbb{Z}_Q$ Berry phase is defined by introducing local-twist parameters
and quantized in $\mathbb{Z}_Q$ ($Q=d+1$) [55].
The definition of the Berry phase for the $d$-dimensional BHT model is the same to that for a square lattice 
model shown in the main manuscript, but the number of twist-parameters is $Q$ (for the detail of the definition,
see Ref. [55]).
Note that the models in [55] are different from the model in [27, 29]
due to the on-site potential $t_1+t_2$, 
but it does not change the Berry phase.
The quantization of the Berry phase is protected by $\mathbb{Z}_Q$ symmetry, which change the annihilation
operator $c_j\rightarrow c_{j+1}, ~c_{Q+1}=c_1$. 
The symmetry is the same to the mirror symmetry for $d=1$ and the three-fold rotational ($C_3$) symmetry for
$d=2$.

\begin{figure}[thb]
 \centering
 \includegraphics[width=0.7\linewidth]{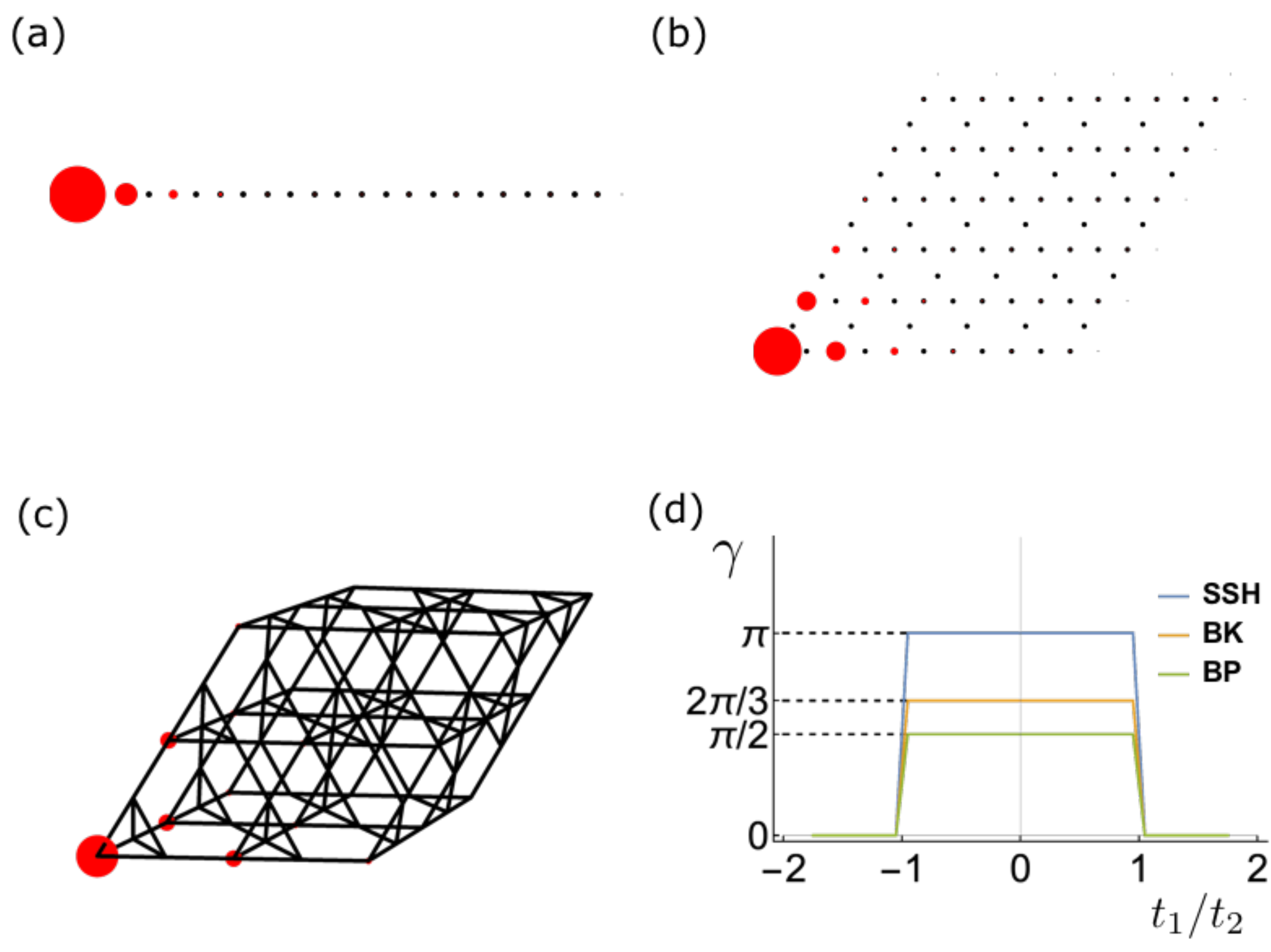}
 \caption{The zero energy corner state of (a) the SSH model, (b) the breathing kagome model (BK) and (c) the
 breathing Pyrochlore (BP) model.
 (d) The $\mathbb{Z}_Q$ Berry phases for the BHT models in $d=1, 2, 3$-dimensions are shown.
 }
 \label{fig:s-fig1}
\end{figure}

The Berry phase of the type-II plaquette for the BHT models are shown in Fig. \ref{fig:s-fig1}(d).
The Berry phases are quantized in $\mathbb{Z}_Q$ for $d = 1, 2, 3$.
The non-trivial Berry phase corresponds with the HOTI phase in which corner states appear 
[Fig. \ref{fig:s-fig1}(a)-(c)].

\subsection{The $Z_Q$ Berry phase for a decoupled cluster}

\begin{figure}[thb]
 \centering
 \includegraphics[width=0.65\linewidth]{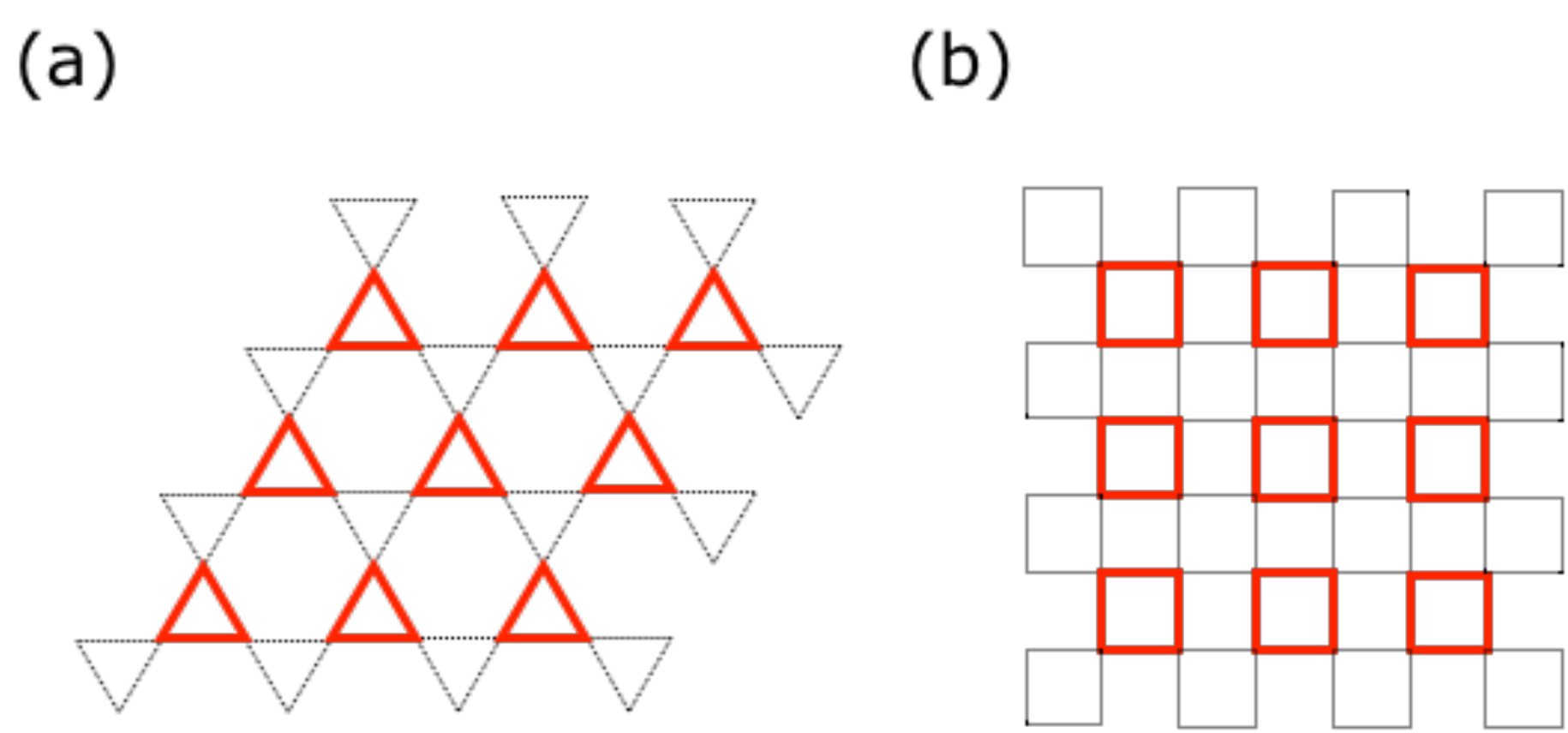}
 \caption{
 Decoupled clusters are colored in red for (a) a kagome lattice and (b) a square lattice.
 }
 \label{fig:s-fig2}
\end{figure}

\begin{figure}[thb]
 \centering
 \includegraphics[width=0.8\linewidth]{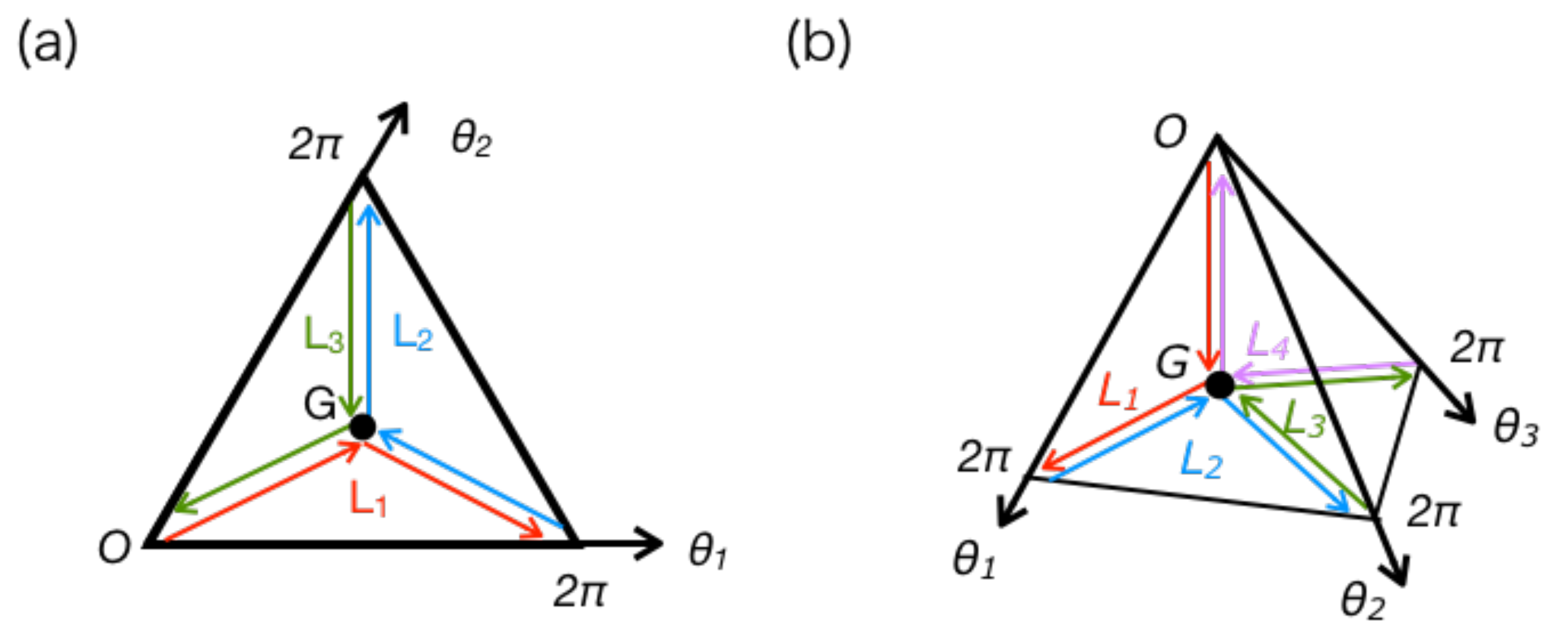}
 \caption{The trajectories of the $Z_Q$ Berry phase for (a) $Q=3$ and (b) $Q=4$ in the parameter space are shown.
 }
 \label{fig:s-fig3}
\end{figure}

The BHT models and the BBH model in the HOTI/HOSPT phases are adiabatically connected to the decoupled clusters [Fig. \ref{fig:s-fig2}].
The Berry phase for the decoupled clusters is the same to that for single cluster because the bond-twists 
only affect to the cluster and the other clusters are not changed.

In this section, we consider single cluster that consists of $Q$ sites and the Hamiltonian of the cluster
has $\mathbb{Z}_Q$ symmetry such that the Hamiltonian is unitary invariant under the cycle 
$c_j \rightarrow c_{j+1},~ c_{Q+1} = c_{1}$ where $c_j (j=1,\cdots, Q)$ is the annihilation operators.

Let $\ket{\Psi}$ be the many-body ground state. We define a unitary operator
\begin{equation}
 U = e ^ { -i \varphi _ { 1 } n _ { 1 } } e ^ { -i \varphi_2 n _ { 2 } } \cdots e ^ { -i \varphi_{Q-1} n _ { Q - 1 } },
\end{equation}
where $\varphi_j = \sum_{i=1}^{j}\theta_i, ~ \varphi_Q = 0$.
The annihilation operators are transformed by the unitary operator $U$ as
\begin{equation}
 U c _ { j }  U ^ { - 1 } = e ^ { i \varphi_j} c _ { j } 
\end{equation}
so the ground state for $\mathcal{H}(\Theta)$ is written as
\begin{equation}
 \label{eq:Psi_theta_is_U_Psi}
 \ket{\Psi(\Theta)} = U\ket{\Psi}.
\end{equation}
Note that the expectation value of the number operators are not changed by $\Theta$ :
\begin{eqnarray}
 \braket{n_j}_\Theta &=& \braket{\Psi(\Theta)| n_j |\Psi(\Theta)} \\
 &=& \braket{\Psi|U^\dagger n_j U|\Psi} \\
 &=& \braket{\Psi|n_j|\Psi} = \braket{n_j}.
\end{eqnarray}

Next, to define the Berry phase, let us introduce trajectories $L_j, (j = 1, \cdots, Q)$ (see Fig. \ref{fig:s-fig3})
\begin{eqnarray}
 L_1 &:& \bm{O} \rightarrow \bm{G} \rightarrow \bm{E}_1 \nonumber \\
 L_2 &:&  \bm{E}_1 \rightarrow \bm{G} \rightarrow  \bm{E}_2 \nonumber \\
 &\cdots& \\
 L_Q &:&  \bm{E}_{Q-1} \rightarrow \bm{G} \rightarrow \bm{O} \nonumber 
\end{eqnarray}
where $\bm{E}_j=2\pi\bm{e}_j$ and \{$\bm{e}_j$\} are unit vectors of the parameter space.
The vector is $\bm{G} = 1/Q\sum_{j=1}^{Q-1}\bm{V}_j$.
We then have
\begin{equation}
\label{eq:total_is_zero_Q}
\sum_{j=1}^Q\gamma_j = 0 ~({\rm mod} ~2\pi).
\end{equation}
due to the cancellation of the trajectories.

The calculation of the Berry phase can be performed explicitly by parametrizing the trajectory.
By using Eq. (\ref{eq:Psi_theta_is_U_Psi}), we have
\begin{eqnarray}
 d\bm{\Theta}\cdot\braket{\Psi(\Theta)|\frac{\partial}{\partial \bm{\Theta}}|\Psi(\Theta)} 
  &=& dt \bra{\Psi(\Theta)}\partial_t\ket{\Psi(\Theta)} \nonumber \\
 &=& -i dt\frac{\partial \theta_1}{\partial t}(\braket{n_1} + \braket{n_2} + \cdots +  
  \braket{n_{Q-1}}) \nonumber \\ 
 &~& - i dt\frac{\partial \theta_2}{\partial t}(\braket{n_2} + \cdots + \braket{n_{Q-1}}) 
  \nonumber \\ 
 &~& - \cdots - i dt\frac{\partial \theta_{Q-1}}{\partial t}\braket{n_{Q-1}},
\end{eqnarray}
where the real parameter $t$ parametrizes the trajectory.
In the trajectory $L_j ~(j = 1, \cdots, Q)$, only the $\frac{\partial \theta_j}{\partial t}$ is $2\pi$ and
the others $\frac{\partial \theta_{k\neq j}}{\partial t}$ are zero.
So the Berry phases, 
\begin{equation}
\gamma_j = -i\int_{L_j}\bra{\Psi(\Theta)}\partial_t\ket{\Psi(\Theta)}dt,
\end{equation}
are
calculated as
\begin{eqnarray}
 \gamma_1 &=& 2\pi (\braket{n_Q} - N \nonumber) \\
 \gamma_2 &=& 2\pi \braket{n_1} \nonumber \\
 &\cdots& \nonumber \\
 \gamma_Q &=& 2\pi \braket{n_{Q-1}}. \label{eq:berry}
\end{eqnarray}
Here, the total number of particles in the cluster, $N = \sum_{j=1}^Q \braket{n_j}$, is an integer, 
so $\gamma_1 = \braket{n_Q} ~({\rm mod} ~2\pi)$. 

Because of the $C_Q$ symmetry, the density of the particle is uniform
\begin{equation}
 \braket{n_1} = \cdots = \braket{n_Q} = N/Q \equiv \nu. \label{eq:berry2} 
\end{equation}
Here $\nu$ is a filling of electrons in a unit cell.
Combining Eqs. (\ref{eq:total_is_zero_Q}) and (\ref{eq:berry2}), 
we finally find the Berry phase of the cluster limit is 
$\gamma_1=\cdots=\gamma_Q\equiv \gamma = 2\pi \nu ~{\rm mod} ~2\pi, ~n \in \mathbb{Z}$.

\section*{BBH model with NNN hopping}
In this section 
we give an exact corner states of the BBH model with NNN hopping.
The tight-binding model is shown in Fig. 2 in the main text. 

First we consider a lattice size of $(2L_x+1) \times (2L_y+1)$. 
In this case, an exact zero energy corner state is obtained:
\begin{equation}
\label{eq:exact_zero}
 \ket{\psi} = \frac{1}{N}\sum_{m,n} r^{m+n}\ket{2m+1,2n+1}.
\end{equation}
Here, $N$ is a normalization factor, $r=-t_2/t_1$ and $\ket{i,j}$ is a basis localized at position $(i, j)$.
One can show that the state is zero energy eigenstate, satisfying $\mathcal{H}\ket{\psi} = 0$.
When $|t_1| < |t_2|$, the corner state is localized at $(1, 1)$
with the localization length $-1/\rm{log}|r|$.

\begin{figure}[thb]
 \centering
 \includegraphics[width=0.4\linewidth]{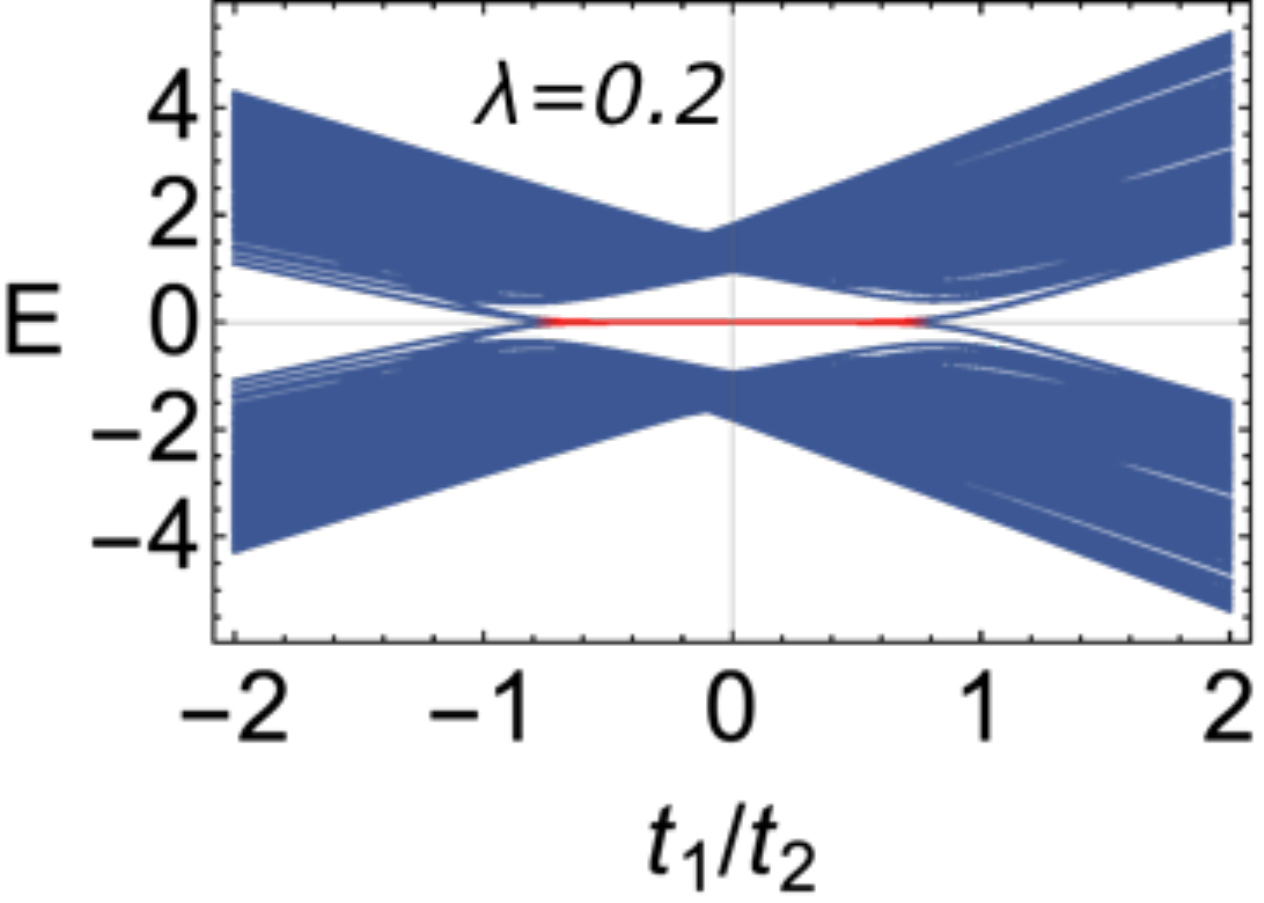}
 \caption{The single particle energy against $t_1/t_2$ with fixed $\lambda=0.2$. 
 The red line denotes the corner localized states. 
 The phase transition points are $t_1/t_2 = \pm 1$.
 }
 \label{fig:s-fig4}
\end{figure}

If the lattice size for both $x$ and $y$ axes are odd, an exact zero energy state (Eq. (\ref{eq:exact_zero})) 
is obtained. For the even case, there are four corner states and the energies of them are not exactly 
zero in a finite system [Fig. \ref{fig:s-fig4}]. 
But when the system size is large enough, 
the exact zero energy state in Eq. (\ref{eq:exact_zero}) gives a good approximation to the corner states 
because the exact zero energy state exponentially decays for distance from a corner and the four corner states
are orthogonal each other, except the phase transition points.

\end{document}